\documentclass[twocolumn]{jpsj3}
\usepackage{bm}
\usepackage{times}
\usepackage{color}

\begin{document}

\title{
Density Modulations Associated with the Dynamical Instability in the Bose-Hubbard Model}

\author{
Rui Asaoka$^{1}$\thanks{E-mail address: asaoka@olive.apph.tohoku.ac.jp}, 
Hiroki Tsuchiura$^{1}$,
Makoto Yamashita$^{2,3}$, and 
Yuta Toga$^{1}$
}
%==============================================================================

\inst{
$^{1}$\address{Department of Applied Physics, Tohoku University, Sendai 980-8579, Japan} \\
$^{2}$\address{NTT Basic Research Laboratories, NTT Corporation, Atsugi, Kanagawa 243-0198, Japan } \\
$^{3}$\address{Japan Science and Technology Agency, CREST, Chiyoda, Tokyo 102-0075, Japan}\\
}

\abst{We analyze the non-equilibrium quantum dynamics of a Bose-Einstein condensate that flows in an optical lattice on the basis of the Bose-Hubbard model. 
The time evolution of a condensate calculated by the dynamical Gutzwiller approximation has clarified that density modulations appear as a precursor phenomenon of the dynamical instability. 
Furthermore, the principal mode of modulations strongly depends on both the interparticle interaction strength and the rate of momentum acceleration. 
We show that these features of density modulations are well explained by the stability phase diagram obtained on the basis of the Bogoliubov theory. 
}

\maketitle
\section{Introduction}
Ultracold atoms in optical lattices shed light on the non-equilibrium quantum dynamics caused by many-body effects or lattice periodicity. Many intriguing phenomena that are hardly observed in conventional solid-state systems have already been reported: Bloch oscillations, Landau-Zener tunneling\cite{anderson,morsch,christ}, resonantly enhanced tunneling\cite{sias}, and solitons at the edge of the Brillouin zone\cite{eiermann}.

A series of experiments have further demonstrated that a Bose-Einstein condensate (BEC) in a periodic potential has a critical momentum where the superfluidity becomes unstable by measuring the center-of-mass oscillations of a BEC or the decay of superfluid flows in a moving lattice\cite{burger,cataliotti,fertig,fallani,cristiani,mun,ferris}. As has been indicated theoretically\cite{wu_niu01,smerzi,wu_niu03,modugno,altman,polkov,danshita,hui}, this phenomenon is identified as the dynamical instability that occurs in the presence of lattice potential, and also without any energy dissipation in contrast to the well-known Landau instability of a superfluid. It is known that, when a BEC is dynamically unstable, an arbitrary small density fluctuation of the original BEC grows exponentially in time, leading to a catastrophic decay of superfluid flow.
 
Fallani {\it et al.}\cite{fallani} observed the nontrivial complex density profiles of atoms after the collapse of a BEC in a shallow one-dimensional (1D) moving optical lattice, which suggests that the dynamical instability causes a certain modulational instability of superfluid density. In this experiment, the system stays in a weakly correlated regime, and the filling of lattice sites is more than several hundred according to the weak confinement in the plane that is perpendicular to the lattice direction. Analyses based on the Gross-Pitaevskii (GP) equation have elucidated that the density modulation is formed as a result of the complicated mode mixing in momentum space\cite{nesi,sarlo}. 
On the other hand, Mun {\it et al.}\cite{mun} experimentally determined the phase boundary of dynamical instability in a wide range of interactions using a BEC in a 3D moving optical lattice. 
Here, the system is described using the Bose-Hubbard (BH) model at low fillings and far from that observed in the experiment by Fallani {\it et al.} It was demonstrated that the measured results are in good agreement with the stability phase diagram of the BH model obtained by the dynamical Gutzwiller approximation\cite{altman,polkov}. However, the existence of density modulations in the vicinity of the dynamical instability remains to be clarified both experimentally and theoretically.

In this work, we analyze the density modulation of a BEC associated with the dynamical instability in the BH model by widely changing interaction strength. The numerical simulations on the basis of the dynamical Gutzwiller approximation have clarified that the density modulation  appears as a precursor of the dynamical instability, having the characteristic wavelength that depends on the interparticle interaction strength and the acceleration rate of the condensate. Moreover, we show that these density modulations can be understood as a manifestation of unstable excitation modes in the Bogoliubov theory.

\section{Model and Methods}
The Hamiltonian of the BH model is given as
%------------------------------------------------------
\begin{eqnarray}
 {\cal H} = -t\sum_{\langle i,j\rangle} 
\left( \hat{a}_{i}^{\dagger}\hat{a}_{j} 
     + \hat{a}_{j}^{\dagger}\hat{a}_{i} \right)
           + \frac{U}{2}\sum_{i}\hat{n}_{i}(\hat{n}_{i}-1)   ,
\end{eqnarray}
%------------------------------------------------------
\noindent
where $\hat{a}_{i}$ ($\hat{a}_{i}^{\dagger}$) is the annihilation (creation) operator of a boson at the site $i$ and $\hat n_{i} = \sum \hat{a}_{i}^{\dagger}\hat{a}_{i}$ is the number operator.  
Here, $\langle i,j\rangle$ in the summation denotes the pairs of nearest neighbors, $t$ ($> 0$) is the single-particle hopping amplitude, and $U$ ($> 0$) is the on-site repulsive interaction.
In the weakly correlated regime ($t\gg U$), the dynamics of a BEC is well described by the GP equation, and the dynamical instability has been first predicted within this framework \cite{wu_niu01,wu_niu03,smerzi,modugno}. However, the GP equation fails in the strongly correlated regime. The dynamical Gutzwiller approximation\cite{altman,polkov} was then developed as an alternative powerful tool that can be applied to the region near the Mott-insulator transition point. This mean-field method is relatively accurate in the higher dimension (i.e., 2D or 3D) where quantum fluctuations are suppressed, while it becomes rather inaccurate in the 1D system because quantum tunneling from metastable states, which is ignored by this theory, plays an essential role in such a low-dimension system.\cite{polkov,mckay,danshita2,montagero}

We investigate the dynamical instability and the associated density modulation of the BEC in a wide range of interaction strengths. We employ the dynamical Gutzwiller approximation based on the following 
variational wave function:
%------------------------------------------------------
\begin{equation}
|\Psi_{\mathrm G}\rangle = \prod_{i}\left[ \sum_{n_{i}=0}^{\infty} f_{i}(n_{i})  
                                               | n_{i} \rangle \right]  ,
\end{equation}
%------------------------------------------------------
\noindent
where $n_{i}$ is the site occupation and $\left\{f_{i}\right\}$ represents a set of Gutzwiller parameters that depend on time. In our calculations, we focus on the unit filling $n=1$ 
and limit the maximum of $n_i$ to $n_{{\rm max}}=5$ to reduce computational tasks. It has been confirmed that this truncation does not affect the present results. 
For instance, the probability that more than six atoms exist at each site was calculated to be less than 0.01\% even in the case where the interaction strength is smallest, i.e., $U/U_{c}=0.2$ and the superfluid flow is absent.

The time-dependent variational principle leads to the equations of motion for the Gutzwiller parameters\cite{polkov}:
%------------------------------------------------------
\begin{eqnarray}
\nonumber
i\dot{f}_{i}(n_{i}) &&= \frac{U}{2}n_{i}(n_{i}-1)f_{i}(n_{i})\\
\nonumber
&&-tz\left\{\sqrt{n_{i}}\,f_{i}(n_{i}-1)\psi _{i}+\sqrt{n_{i}+1}\,f_{i}(n_{i}+1)\psi _{i}^{*}\right\}, 
\\
\label{eom}
\end{eqnarray}
%------------------------------------------------------
\noindent
where $\nonumber \psi _{i} = \frac{1}{z}\sum_{j}\langle\Psi_{\mathrm G}|\hat{a}_{j}e^{ip(x_{j}-x_{i})}|\Psi_{\mathrm G}\rangle$ and the summation of $j$ runs over all the nearest neighbors around the site $i$. Here, the phase factor $p$ is the momentum of flow and  $x_{i}$ is the position of the site $i$. The quantity $z$ corresponds to the number of adjacent sites given by $z=2\times d$ ($d$: dimension). 
For a uniform system, we generally neglect the  of site dependence of $f_i(n_i)$ in Eq.\,(\ref{eom}) and reduce the number of Gutzwiller parameters to $(n_{\rm max}+1)$ in any dimensions.
However, as we will see later, the density modulations are generated only in the direction of superfluid flow, which makes the system inhomogeneous in this direction. 
To analyze the characteristic wavelengths of these density modulations, we should carefully consider the site dependence of Gutzwiller parameters 
and deal with a sufficiently large number of lattice sites in the direction of superfluid flow \cite{altman,polkov}.
On the other hand, the system maintains its homogeneity in the perpendicular direction of the flow even after the dynamical instability occurs.

In this work, we investigate the 2D BH model in units of  $160 \times 2$ lattice sites by imposing the periodic boundary condition on the system.  
The number of Gutzwiller parameters in our calculations is therefore $160 \times 2 \times 6=1\,920$.
We determine these parameters by numerically solving Eq.\,\,(\ref{eom}) in a self-consistent manner under the condition that the momentum is increased almost adiabatically, $p=\alpha t$ ($t$ : time) with a tiny coefficient $\alpha$.

\section{Results and Discussion}
\begin{figure}
\begin{center}
\includegraphics[width=8.0cm]{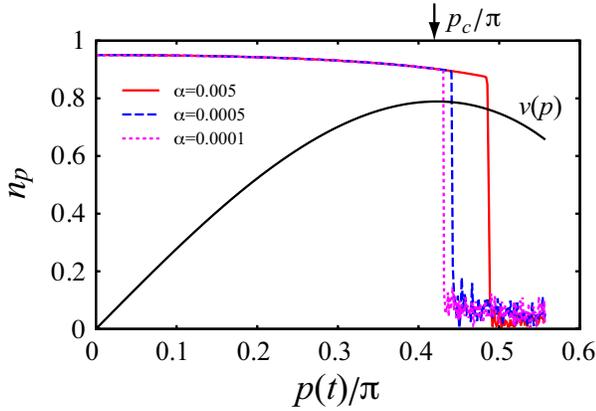}
\caption{(Color online)
Condensate fraction $n_{p}$ as a function of the momentum $p$ that is increased proportionally to time $t$ such as $p(t)=\alpha t$. 
We assume the BH model with the interaction strength $U/U_{c}=0.2$. The group velocity of condensates, $v(p)$, becomes largest at the critical momentum $p_{c}$ indicated by the arrow. Condensate fractions abruptly decrease after $p$ exceeds $p_{c}$, reflecting the collapse of condensates.
}
\label{time}
\end{center}
\end{figure}

Figure \ref{time} shows the evolution of the condensate fraction $n_{p}$ when the momentum $p$ is increased gradually in the BH model. Here, the condensate fraction is defined as the population of the Bloch state with the momentum $p$. We choose three different $\alpha$ values and set the interaction strength at $U/U_{c}=0.2$, where $U_{c}$ is the critical value at the Mott-insulator transition point.
As shown in Fig.\,\ref{time}, $n_{p}$ abruptly decreases at a certain momentum $p_{c}'$ for each $\alpha$, which reflects that the condensate collapses via dynamical instability. We find that this collapse momentum $p_{c}'$ becomes larger as $\alpha$ is increased.
On the other hand, a critical momentum of superfluid flow, $p_{c}$, a superfluid flow can dynamically collapse at $p > p_{c}$, is determined from the (non-dimensional) group velocity given by $v(p)=\rho(p)\sin(p)$, where $\rho(p) $ corresponds to the density of a superfluid that flows {\it steadily} with the momentum $p$. The periodicity of $v(p)$ reflects the structure of the lowest Bloch band in an optical lattice. Furthermore, $\rho(p) (\propto t'/U)$ is a monotonically decreasing function of $p$ according to the effective hopping amplitude $t'= t(d+\cos(p)-1)/d$. The group velocity $v(p)$ is therefore expected to become a maximum at a certain momentum $p=p_{c}(<\pi/2)$.  
We obtained this $v(p)$ separately by evaluating the superfluid density variationally on the basis of the conventional (i.e., not the dynamical) Gutzwiller approximation in a way such that $\rho(p)=|\langle \hat a_{p}\rangle|^{2}$, where $\hat a_p$ denotes the annihilation operator of a boson having momentum $p$\cite{verification}. The results are also shown in Fig.\,\ref{time}. As has been expected, $v(p)$ reaches its maximum at the momentum $p_{c}/\pi=0.42$ indicated by the arrow.
It is known that the effective mass of a superfluid is inversely proportional to the derivative $dv/dp$. Thus, for $p > p_{c}$, the effective mass has a negative sign and the superfluid flow becomes unstable\cite{polkov}. 
The difference between the points $p_{c}$ and $p_{c}'$ corresponds to the delay for the full development of the unstable mode. We see in Fig.\,\ref{time} that $p_{c}'$ approaches $p_{c}$ as $\alpha$ is decreased. 
This difference, however, always exists in the experiments since complete adiabaticity is impractical experimentally. 

%=====================================================================
%=====================================================================

\begin{figure}
\begin{center}
\includegraphics[clip,width=8.0cm]{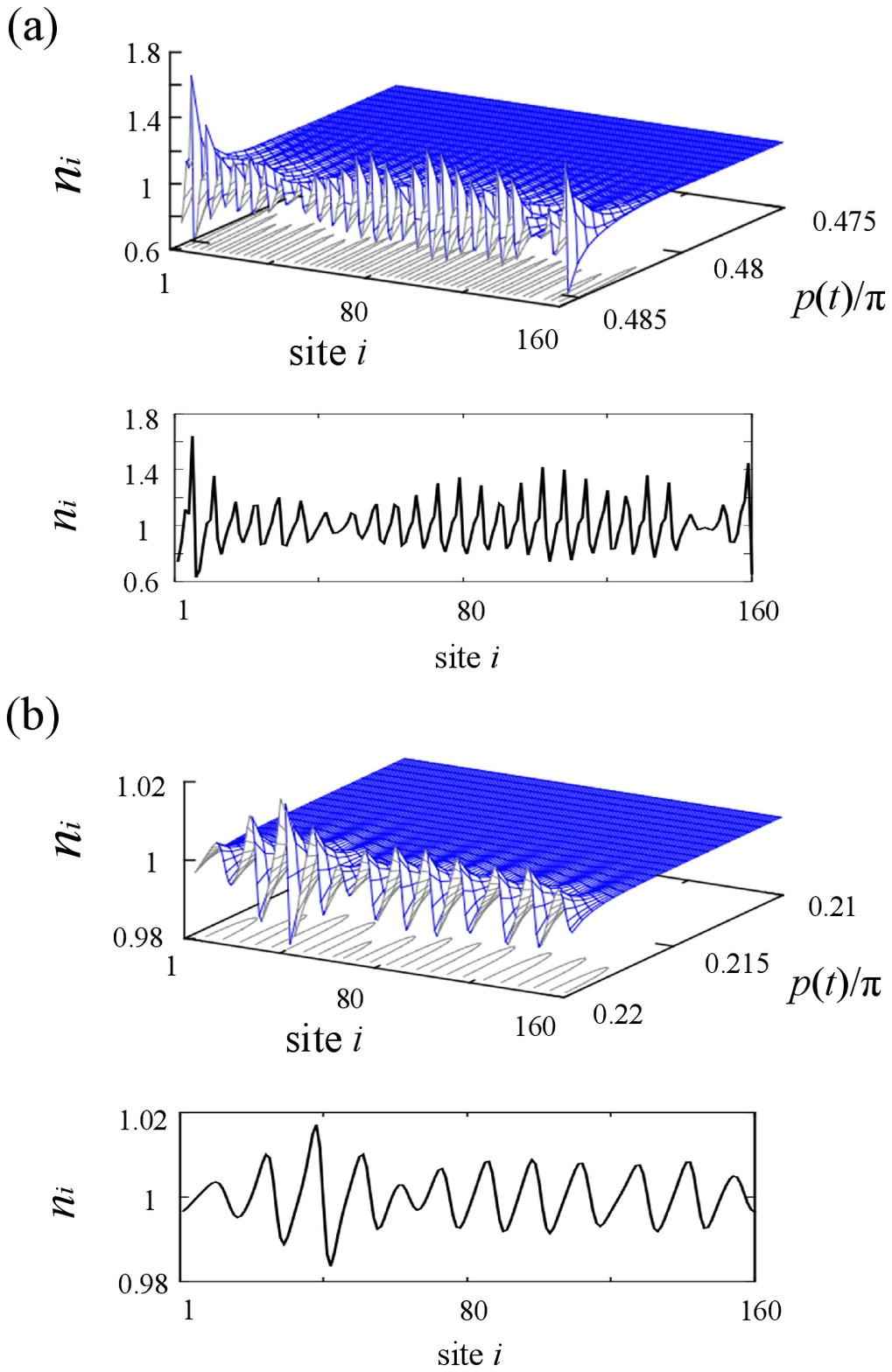}
\caption{(Color online)
Time evolution of density distribution over the sites as $p(t)$ approaches the collapse momentum $p_c'$. We choose the different interaction strengths in the BH model: (a) $U/U_{c}=0.2$ and (b) $U/U_{c}=0.8$. The rate of momentum acceleration is set at $\alpha=0.005$. The bottom figures in (a) and (b) exhibit cross-sectional views of density distributions at $p/\pi=0.486$ and $0.22$, respectively. The critical momentum and collapse momentum are (a) $(p_{c}/\pi, \,p_{c}'/\pi)=(0.423, 0.486)$ and (b) $(0.178, 0.22)$. The sites in the figure are numbered in the direction of superfluid flow.
}
\label{wave}
\end{center}
\end{figure}

Next, we discuss the dynamics of the condensate in real space. In Fig.\,\ref{wave}, we show the time evolution of the density distribution over the lattice sites around the collapse momentum $p_{c}'$. 
The parameters used in the calculations are (a) $U/U_{c}=0.2$ and (b) $U/U_{c}=0.8$, and the critical momenta and collapse momenta are (a) $p_{c}/\pi=0.423$, $p_{c}'/\pi=0.486$, and (b) $p_{c}/\pi=0.178$, $p_{c}'/\pi=0.22$, respectively. Here, the rate of momentum acceleration is assumed to be $\alpha=0.005$. From Fig.\,\ref{wave}, one can see that the density modulation occurs as a precursor of the dynamical instability and grows gradually only in the direction parallel to the flow. 
Furthermore, the cross sections in Fig.\,\ref{wave} clarify that the wavelength of density modulation becomes longer for larger $U$, which suggests that the principal mode of the collective excitation associated with the dynamical instability depends on the interaction strength. We examine these results by changing the unit size of lattice sites in our Gutzwiller analysis and find that both $p_{c}'$ and the principal-mode wavelength of density modulation remain unchanged when the unit size in the flow direction exceeds ten sites.

%=====================================================================
%=====================================================================

\begin{figure}
\begin{center}
\includegraphics[clip,width=8.0cm]{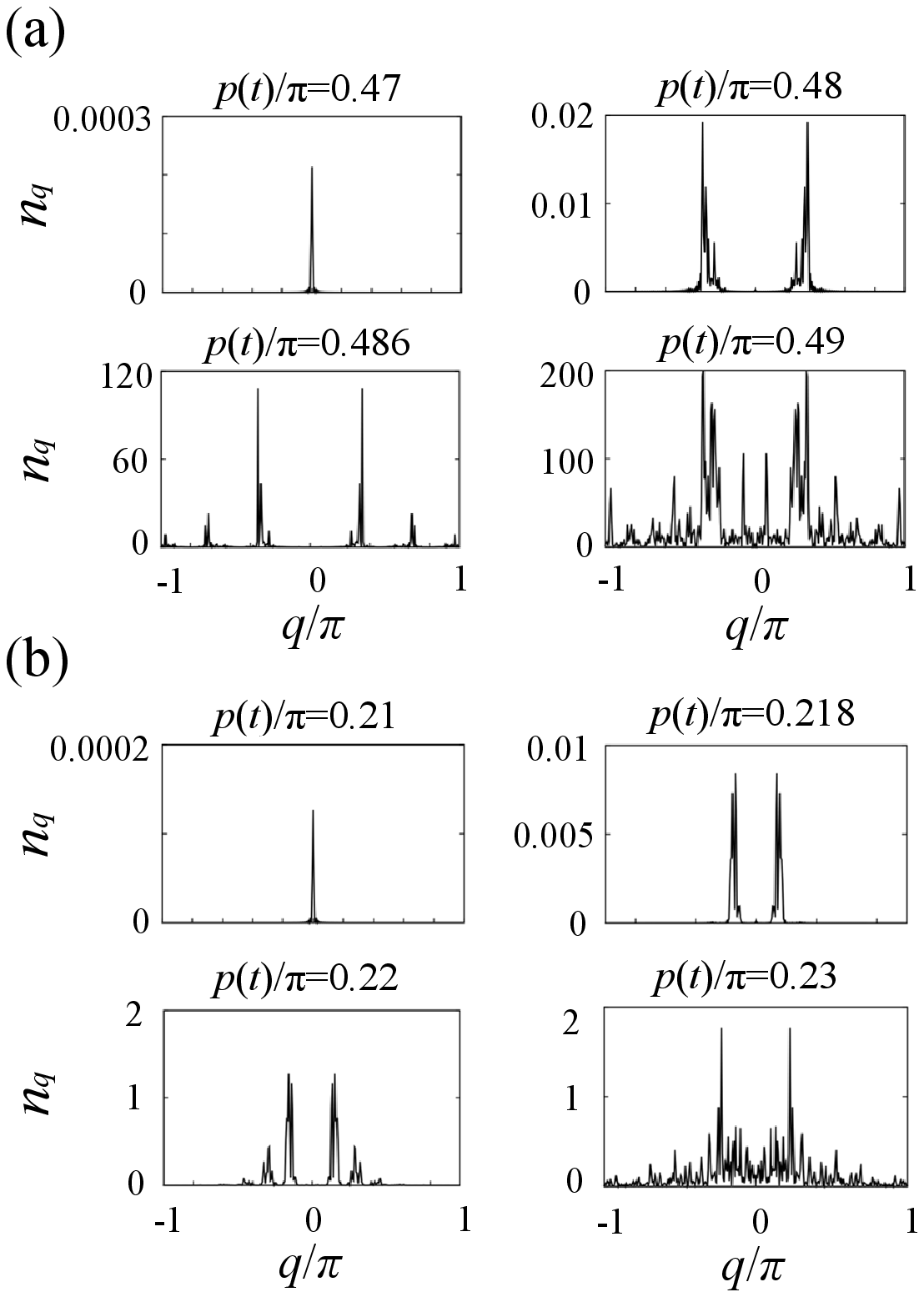}
\caption{
Fourier transform of $\delta n_i (=n_{i}-n$) corresponding to Fig.\,\ref{wave} and that after the collapse of the condensate. (a) $U/U_{c}=0.2$ and (b) $U/U_{c}=0.8$. $q$ is the wave number of density modulation. Main peaks appear at $q/\pi\sim 0.32$ in (a) and $q/\pi\sim 0.16$ in (b).
}
\label{fourier}
\end{center}
\end{figure}

To study the above density modulations more quantitatively, we show in Fig.\,\ref{fourier} the time evolution of the Fourier transform of $\delta n_i (=n_{i}-n$) for $U/U_{c}=0.2$ and $0.8$ corresponding to Fig.\,\ref{wave}. We see the common features in both cases: first, the main peak emerges at a certain wave number (figures in the upper right); then, the side peaks grow at wave numbers that are integral multiples of the main peak (lower left); finally, many peaks develop simultaneously at various wave numbers as a signature of dynamical instability (lower right). Here, the main peaks of $n_{q}$ appearing at the early stage correspond to the principal modes of density modulations shown in Fig.\,\ref{wave} and their wave numbers are calculated to be $q/\pi \sim 0.32$ for $U/U_{c}=0.2$ and $q/\pi \sim 0.16$ for $U/U_{c}=0.8$. We further verify that these values become smaller as $\alpha$ is decreased. The principal mode therefore depends on the strength of the interparticle interaction $U$ and the rate of momentum acceleration $\alpha$. On the other hand, the growth of side peaks indicates that additional density modulations are induced by the mean-field potential originated from the density distribution of principal mode. Moreover, from Fig.\,\ref{fourier}, the principal modes of density modulations strongly survive after the condensate collapses. Note that these simple features are in clear contrast to the complicated dynamics of mode mixing obtained using the GP equation in Ref.\,22.

%=====================================================================
%=====================================================================

We can understand the density modulations associated with the dynamical instability in Fig.\,\ref{fourier} as a manifestation of unstable excitation modes in the Bogoliubov theory.  Here, we analyze 
the stability phase diagram within a truncated Hilbert space. We employ the following Gutzwiller variational wavefunction\cite{auer}:
%-----------------------------------------------------
\begin{eqnarray}
\nonumber
|\mathrm G\rangle &&=\prod_{i}\left[\cos\frac{\theta_{i}}{2}|n\rangle_{i}+e^{i\eta_{i}}\sin\frac{\theta_{i}}{2}\left(\cos\frac{\chi_{i}}{2}e^{-i\varphi_{i}}|n-1\rangle_{i}\right.\right.\\
&&\hspace{10mm}+\left.\left.\sin\frac{\chi_{i}}{2}e^{i\varphi_{i}}|n+1\rangle_{i}\right)\right]    ,
\label{gutz2}
\end{eqnarray}
%-----------------------------------------------------
\noindent
where the phase $\varphi_{i}$ is given by the product of momentum and position: $\varphi_{i}={\bf p}\cdot{\bf x}_{i}$.  The value $\theta_i=0$ corresponds to the Mott-insulator phase, while $\theta_i > 0$ describes the superfluid phase. Owing to the truncation of the states, this variational wavefunction is generally applicable only in the region around the Mott-insulator transition point. However, in the present analysis, the effective hopping amplitude is given by $t'\sim t\cos(p)$ and decreases as the condensate momentum $p$ is increased. For instance, the probability that more than three atoms exist at each site is calculated to be smaller than 0.9\% at the point of the critical momentum for $U/U_{c}=0.2$. The function $|\mathrm G\rangle $ therefore becomes valid even in the weakly correlated regime when $t'$ is sufficiently small. The variational parameters for the stationary state are determined so as to minimize the energy per site, and we obtain $\chi=\frac{\theta}{2}$, $\eta=0$, and $\cos\theta=\frac{U}{(3+2\sqrt{2})tz\gamma_{{\bf p}}}$ with $\gamma_{{\bf p}}=\frac{\cos p+1}{2}$ for $n=1$ in the 2D system. Note that we deal with the homogeneous system, and the suffix $i$ is omitted from all the variational parameters for simplicity.

We discuss the collective excitation modes around this stationary state in the framework of the Bogoliubov theory. For a given $U/t$ and ${\bf p}$, the excitation modes are given by the Bogoliubov transformation, which corresponds to the evaluation of the eigenvalues $\epsilon$ for the matrix $\sigma\cal{M}$ in Ref.\,20 (spinless case), where $\sigma \equiv {\rm diag} (1,-1,1,-1) $. It is straightforward to obtain the analytical form
%------------------------------------------------------
\begin{eqnarray}
\nonumber
&& \epsilon ^{2} = \frac{1}{2}\Big[A^{2}+B^{2}-C^{2}-D^{2}  \pm \big\{(A^{2}-B^{2}-C^{2}+D^{2})^{2}\\
&&\hspace{20mm}+16(A+C)(B+D)E^{2}\big\}^{\frac{1}{2}} \Big] ,
\label{eagen}
\end{eqnarray}
%------------------------------------------------------
where
%------------------------------------------------------
\begin{eqnarray}
\nonumber
&&A=2\gamma_{{\bf p}}-\cos^{2}\theta\gamma_{+}, \ B=(2\gamma_{{\bf p}}-\gamma_{+})\cos^{2}\frac{\theta}{2},\\
&&C=\cos^{2}\theta\gamma_{+}, \ D=\cos^{2}\frac{\theta}{2}\gamma_{+}, \ E=\cos\theta\cos\frac{\theta}{2}\gamma_{-}.
\end{eqnarray}
%------------------------------------------------------
Here, $\gamma_{\pm}=\frac{1}{2}(\gamma_{{\bf k}+{\bf p}}\pm \gamma_{{\bf k}-{\bf p}})$ and $k \in [-\pi , \pi]$. The dynamical instability occurs when at least one of the $\epsilon$ values has an imaginary part. From Eq.\,(\ref{eagen}), we finally derive the instability condition
%------------------------------------------------------
\begin{eqnarray}
\nonumber
\cos^{2}k&-&\frac{2\cos p}{u^{2}}\left(\gamma_{{\bf p}}^{3}+u^{2}\gamma_{{\bf p}}-u^{2}\right)\cos k
\\
\nonumber
&&+\frac{1}{u^{2}}\left(4\gamma_{{\bf p}}^{4}-2\gamma_{{\bf p}}^{3}+4u^{2}\gamma_{{\bf p}}^{2}-6u^{2}\gamma_{{\bf p}}+u^{2}\right) < 0    ,\\
\label{insta}
\end{eqnarray}
%------------------------------------------------------
\noindent
where $u\equiv \frac{U}{4(3+2\sqrt{2})}$ for $t=1$ and $z=4$.

\begin{figure}
\begin{center}
\includegraphics[clip,width=8.0cm]{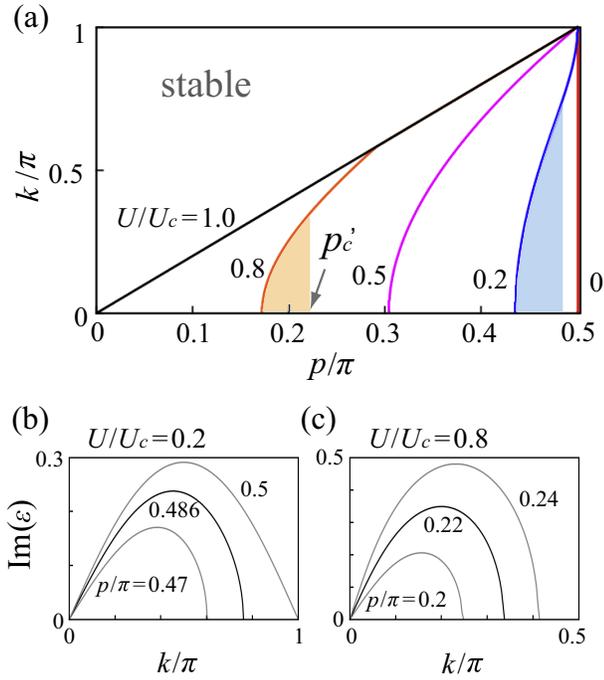}
\caption{(Color online)
(a) Stability phase diagram of superfluid flow for several interaction strengths. $p$ is the momentum of condensates and $k$ is the wave number of excitation modes. A condensate is stable in the left-side area (unstable in the right-side area) with respect to the boundary for each interaction strength. The point at $k=0$ on each boundary line gives the critical momentum $p_{c}$.
The shaded areas represent the unstable modes related to the density modulations in Figs.\,\ref{wave} and \ref{fourier}. 
 (b) and (c) Imaginary part of eigenvalues $\epsilon$ as a function of wave number $k$. We choose the following parameters: (b) $U/U_{c}=0.2$, $p/\pi=0.47$, $0.486$, and $0.5$; (c) $U/U_{c}=0.8$, $p/\pi=0.2$, $0.22$, and $0.24$. The thick lines correspond to the case where $p=p_c'$. We further assume $n=1$ and $d=2$ in (a), (b), and (c). 
}
\label{diag}
\end{center}
\end{figure}

In Fig.\,\ref{diag}(a), the stability boundaries obtained from Eq.\,(\ref{insta}) are plotted for several $U$ values. 
A condensate is unstable in the area on the right side of the boundary for each interaction strength, and this unstable area extends as $U$ increases. 
On the boundary line, the $k=0$ mode always gives the smallest momentum that induces the dynamical instability regardless of interaction strength. 
The $k=0$ mode first makes a condensate unstable if the momentum $p$ is increased fully adiabatically.
The point at $k=0$ on each boundary line gives the critical momentum $p_{c}$ for the corresponding $U$. 
On the other hand, as seen in Fig.\,\ref{time}, a condensate collapses at a larger momentum $p_{c}'$ instead of at $p_{c}$ because the rate of momentum  acceleration $\alpha$ is inevitably finite. 
Thus, the density modulations in Figs.\,\ref{wave} and \ref{fourier} can be induced by the unstable modes 
having $k\neq 0$ wave numbers between $p_{c}$ and $p_{c}'$, which are depicted by the shaded areas for $U/U_c=0.2$ and $0.8$ in Fig.\,\ref{diag}(a). 
Next, we elucidate the important properties of these unstable modes by showing the imaginary part of eigenvalues $\epsilon$ in Figs.\,\ref{diag}(b) and \ref{diag}(c). Note that we choose several momentum values at approximately $p_c'$. 
As a common feature, the imaginary part reaches its maximum at a certain wave number, and the maximum point moves toward a larger wave number as the momentum $p$ increases. 
On the other hand,  as a significant dependence on the interaction strength, the wave number at its maximum becomes smaller for a larger $U$.

Here, we briefly discuss the results in Figs.\,\ref{wave} and \ref{fourier} using Fig.\,\ref{diag}. Figure \ref{wave} shows that the density modulation of a condensate appears just before 
$p$ reaches $p_c'$. This suggests that, in Fig.\,\ref{diag}(a), the unstable modes in the vicinity of $p=p_c'$ (i.e., near the right edge in the shaded area) are strongly related to the density modulations. 
Moreover, Fig.\,\ref{fourier} clarifies that the density modulations are well characterized by the wave number of the principal mode: 
$q/\pi\sim 0.32$ for $U/U_c=0.2$ and $q/\pi\sim 0.16$ for $U/U_c=0.8$. 
These $q$ values reasonably are in reasonable agreement with the wave number around the maximum point of the thick line for $p=p_c'$ in Figs.\,\ref{diag}(b) and \ref{diag}(c), respectively. 
The density modulations in Figs.\,\ref{wave} and \ref{fourier} can be finally identified as the unstable excitation modes with the maximum imaginary part of the eigenvalues in the Bogoliubov theory,  
which leads to the largest decay rate of condensates.  
Similarly, we discuss the dependence of density modulation on the rate of momentum acceleration $\alpha$. 
Figure \ref{time} shows that $p_c'$ approaches $p_c$ as $\alpha$ is decreased. From Fig.\,\ref{diag}, 
it can be expected that the wave number of the principal mode $q$ decreases with decreasing $\alpha$, as we have numerically confirmed by the dynamical Gutzwiller approximation.

%=====================================================================
%=====================================================================

\section{Summary}
We have analyzed the density modulation of condensates as a precursor of the dynamical instability in the Bose-Hubbard model. The numerical simulations based on the dynamical Gutzwiller approximation elucidate that the principal mode of the density modulation highly depends on the interaction strength $U$ and the momentum acceleration rate $\alpha$. 
This mode has a wave number $k\neq 0$ owing to the finite $\alpha$ and remains strong even after the collapse of a condensate.  
Furthermore, these features are consistent with the stability phase diagram calculated on the basis of the Bogoliubov theory.
It might be possible to observe the density modulations in the BH model that we have studied in this work by employing the microscopic measurement techniques developed recently\cite{bakr}.
An analysis considering the long-range correlation beyond the present mean-field-approximation is essential for discussing the experimental observation quantitatively.

%%%%%%%%%%%%%%%%%%%%%%%%%%%%%%%%%%%%%%%%%%%%%%%
\begin{acknowledgments}
%%%%%%%%%%%%%%%%%%%%%%%%%%%%%%%%%%%%%%%%%%%%%%%
Some of the numerical computations were carried out at the Yukawa Institute Computer Facility and at the Cyberscience Center, Tohoku University. This work was partly supported by JSPS KAKENHI Grant Number 25287104.
\end{acknowledgments}

%%%%%%%%%%%%%%%%%%%%%%%%%%%%%%%%%%%%%%%%%%%%%%%

\nocite{*}


\begin{thebibliography}{9}
\bibitem{verification}
In Fig.\,\ref{time}, the condensate fraction $n_p(t)$ is calculated under the condition that the momentum $p(t)$ is increased very slowly depending on the small $\alpha$ coefficients. It can be expected that the time evolution of the superfluid state becomes almost steady, and $n_p(t)$ in Fig.\,\ref{time} is nearly equal to $\rho(p)$ until the dynamical instability occurs [i.e., $p(t)< p_c'$]. We introduce the instantaneous group velocity defined by the expression $v(t)= n_p(t) \sin[p(t)]$ and verify that the momentum at the point where this $v(t)$ shows its maximum is in agreement with the critical value $p_{c}$.
\end{thebibliography}
\end{document}